\documentclass[twocolumn
]{ceurart}

\usepackage{listings}
\begin{document}

\copyrightyear{2025}
\copyrightclause{Copyright for this paper by its authors.
  Use permitted under Creative Commons License Attribution 4.0
  International (CC BY 4.0).}

\conference{PatentSemTech’25: 6th Workshop on Patent Text Mining and Semantic Technologies}

\title{Efficient Patent Searching Using Graph Transformers}


\author[1]{Krzysztof Daniell}[
email=krzysztof@iprally.com,
orcid=0009-0006-5959-1804
]

\author[1]{Igor Buzhinsky}[
email=igor@iprally.com,orcid=0000-0003-3713-6051
]

\author[1]{Sebastian Björkqvist}[
email=sebastian@iprally.com,
orcid=0009-0006-9039-8623
]

\address[1]{IPRally Technologies Oy, Helsinki, Finland}


\begin{abstract}
Finding relevant prior art is crucial when deciding whether to file a new patent application or invalidate an existing patent. However, searching for prior art is challenging due to the large number of patent documents and the need for nuanced comparisons to determine novelty. An accurate search engine is therefore invaluable for speeding up the process. We present a Graph Transformer-based dense retrieval method for patent searching where each invention is represented by a graph describing its features and their relationships. Our model processes these invention graphs and is trained using prior art citations from patent office examiners as relevance signals. Using graphs as input significantly improves the computational efficiency of processing long documents, while leveraging examiner citations allows the model to learn domain-specific similarities beyond simple text-based matching. The result is a search engine that emulates how professional patent examiners identify relevant documents. We compare our approach against publicly available text embedding models and show substantial improvements in both prior art retrieval quality and computational efficiency.
\end{abstract}


\maketitle

\section{Introduction}

Companies and individuals can protect their intellectual property by filing a patent application for their inventions. When granted, a patent provides its holder exclusive rights to the invention for a limited time. In exchange, the invention is publicly disclosed. The patent application process can be quite costly and time-consuming, so applicants typically seek a high certainty of it eventually being granted before initiating the application process. Since patents are granted only for novel inventions, conducting a prior art search is important to avoid unnecessary costs and delays. Finding relevant prior art is challenging due to the large number of existing patent documents and the detailed analysis required to distinguish novelty-destroying prior art from merely similar inventions. An effective patent search engine will both speed up the process and improve the results.

In this work, we present an approach for patent searching using graph representations of inventions as input to a Graph Transformer network. The graph of a patent document describes the core of the invention disclosed in that document, namely the features and the relationships between them. This representation condenses the original document, significantly reducing the computational resources required to process the document. The graphs are used as input to a Graph Transformer model trained on patent office examiner citations created during the application process for a patent. This enables the model to capture relationships between similar inventions, even when described using different terms. Furthermore, examiner citations help the model learn domain-specific terminology across different technical domains. The result is a patent search engine that efficiently retrieves relevant novelty-destroying prior art.



\begin{figure*}
  \includegraphics[width=\textwidth]{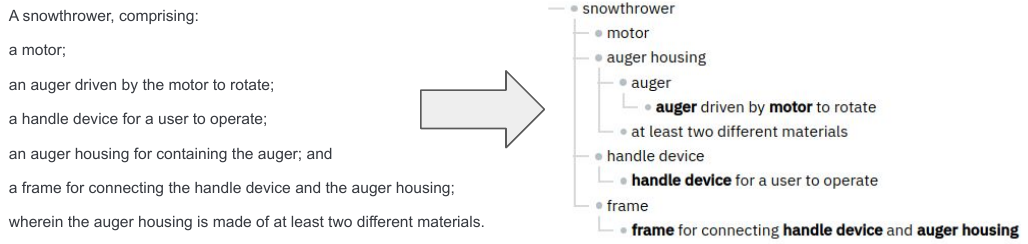}
  \caption{Example patent claim (US20170152638A1) converted into a graph.}
  \label{fig:snowthrower}
\end{figure*}

\section{Background} \label{sec:background}

\subsection{Patent Search Methods}

Traditionally, patent searching has relied on Boolean search, where text-based matching is done using Boolean operators like AND, OR, and NOT \cite{clarke-basics}. Multiple Boolean search tools for patent searching exist, both paid \cite{patbase, patsnap, orbit} and free \cite{google-patents, espacenet, uspto-search}. Performing high-quality Boolean searches often requires multiple iterations and substantial domain expertise to identify the most relevant terminology \cite{ali2024innovating}.

Machine learning methods have been explored for patent searching to address the limitations of Boolean searching. Traditional approaches including TF-IDF \cite{patent-search-tf-idf, d2010clef} and word embeddings \cite{automatic-patent-search} have been applied. More recently, deep learning models such as BERT \cite{ghosh2024paecter, acikalin2022patentsearch, freunekbodmertransformerpatentsearch} and GPT-2 \cite{search-and-reranking} have also been explored. In addition, several commercial machine learning-based patent search tools are available \cite{ambercite, amplified, ipscreener}.

In our previous work \cite{bjorkqvist2023building}, we describe a graph-based patent search engine where a Tree-LSTM model is trained for dense retrieval using patent office examiner citations as relevance signals. This paper extends that work by replacing the Tree-LSTM model with a Graph Transformer and improving the training procedure, as detailed in Section \ref{sec:approach}.

\subsection{Graph Transformers}

Graph Transformers adapt the Transformer architecture to graph-structured data \cite{graph-transformers-survey}. Early models, such as Graphormer \cite{graphormer} and SAN \cite{san}, used full attention over all node pairs, which becomes computationally prohibitive as graph size grows---a critical concern for large patent corpora. Sparse Graph Transformers \cite{sparse-graph-transformer} address this by restricting attention to actual edges or small neighborhoods, drastically reducing overhead while preserving power for long-range dependencies. This is useful for patent searching where long documents revolving around a limited set of distinct features or relationships are typical. Mapping these concepts to nodes, with edges only where genuine conceptual links exist, yields a much sparser graph than a fully connected alternative. 

\subsection{Context-Based Representations}

A recent approach in language modeling is using concept-level representations instead of token-based architectures. Large Concept Models \cite{large-concept-models} can be used to embed entire sentences---or “concepts”---into a high-dimensional, language-agnostic space rather than embedding every token separately. This perspective parallels our approach, where each key textual segment (e.g., a feature of the invention) is treated as a “concept node.” We reduce computational overhead by applying sparse attention to these conceptual links yet still preserve the broad relationships essential for performing patent searches.

\section{Approach} \label{sec:approach}

Our patent search engine consists of two main parts: We first convert each patent document into a graph that describes the invention, as described in Section \ref{sec:graphs}. We then use a Graph Transformer model, described in Section \ref{sec:model-architecture}, to embed the graph into a vector space for dense retrieval. The Graph Transformer model is trained for patent searching using patent examiner citations as relevance signals, as described in Sections \ref{sec:training-data} and \ref{sec:training-procedure}.

The main differences to our previous work \cite{bjorkqvist2023building} are replacing the Tree-LSTM model with a Graph Transformer and enhancements to the training procedure.

\subsection{Graphs}\label{sec:graphs}
We convert each patent document into a graph to capture its core features and relationships while avoiding the overhead of full-text processing. An example of a graph is seen in Figure \ref{fig:snowthrower}. Each node corresponds to a key feature of the invention (e.g., \textit{snowthrower} or \textit{motor}) or a text snippet indicating a relationship (e.g., \textit{frame for connecting handle device and auger housing}). These relationships include hierarchical ones---such as part-of (meronym) or example-of (hyponym)---and functional ones, linking specific features to describe how they interoperate. By focusing on the essential technical structure, the graph becomes much smaller than the raw text while still conveying the essential features of the invention. This method aligns with how professional examiners perform novelty evaluations: they isolate core features and examine how they interrelate. 

The details of how the graphs are created are described in our previous work \cite{bjorkqvist2023building}. In short, we first detect the features of the invention by doing a linguistic analysis with a natural language processing model. To find the relationships between the features, we use a set of hand-crafted rules designed to mirror the way patent professionals identify core inventive concepts and their interconnections from patent text. These rules detect terms describing relationships (e.g., comprising, connecting, containing) and use the output from the linguistic analysis phase to match the relationships with the correct features.

For each patent document, we create three different graphs: One containing only the first independent claim (later referred to as \textit{first claim graph}), another containing all the claims (\textit{all claims graph}), and one containing both the claims and the full description of the document (\textit{description graph}).

\subsection{Model Architecture} \label{sec:model-architecture}

\begin{figure}
    \centering
    \includegraphics[width=0.94\linewidth]{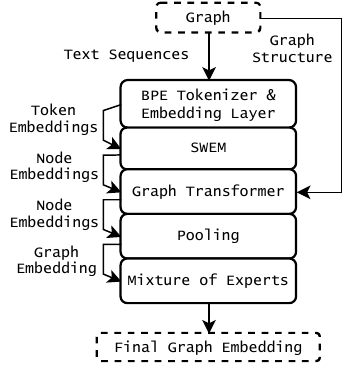}
    \caption{Proposed model architecture. We create initial node embeddings by using a BPE tokenizer and SWEM. The Graph Transformer layers refine these embeddings. We create a graph embedding as a weighted sum of all node embeddings and reduce the dimensionality of the graph embedding using a MoE layer.}
    \label{fig:model-flow}
\end{figure}

Our approach embeds each invention graph into a vector space for dense retrieval. Figure \ref{fig:model-flow} illustrates four key steps, which we describe in detail below.

\subsubsection{Node Embedding Initialization}\label{sec:embedding-layer}
Each token sequence (e.g. sentence or phrase) is tokenized using a BPE tokenizer \cite{bpe-tokenizer} trained on patent documents. We pre-train the token embeddings using FastText \cite{fasttext} to improve the convergence speed of the model training. For each token sequence, we apply a Simple Word-Embedding-based Model (SWEM) \cite{swem}, which computes mean and max pooled embeddings combined with a linear projection. This approach captures contextual information with a lower overhead than LSTMs or CNNs. 

\subsubsection{Graph Transformer Layers}
Next, we feed the node embeddings into a Graph Transformer consisting of several layers that operate on the invention graph. We use Query-Key normalization \cite{qk-norm} to stabilize training and reduce gradient variance. We also adopt a Pre-LayerNorm \cite{pre-layernorm} architecture for consistent gradient flow. Each feed-forward sublayer uses Gated Linear Units with a GELU activation (GEGLU) \cite{geglu}. 
We define adjacency according to the edges in the invention graph, ensuring that sparse attention focuses only on closely related nodes. This lowers computational costs while preserving crucial inter-node relationships \cite{sparse-graph-transformer}.

\subsubsection{Pooling}
After the Graph Transformer layers refine each node representation, we assign each node a learned importance weight. We then compute a graph embedding using a weighted sum of all node embeddings, emphasizing the nodes most relevant to downstream tasks.

\subsubsection{Dimensionality Reduction Layer}\label{sec:dimensionality-reduction}
Lastly, we use a densely gated Mixture of Experts (MoE) \cite{moe} network to project the graph embedding to a lower dimension.

\subsection{Training Data} \label{sec:training-data}

The relevance signals used for training our model are citations extracted from patent examiner reports of patent applications. This approach ensures high-quality, expert-curated training data, as examiner citations highlight legally and technically relevant prior art---what we wish our search engine to place on top of search results. 

We use the following citation categories: novelty-destroying prior art (labeled \textit{X} by the EPO \cite{epo-examination-guidelines}), relevant prior art that does not invalidate novelty (\textit{A citation}), and documents that show that the invention follows in an obvious way as a combination of existing inventions (\textit{Y citation}). 
We include citations from more than 40 jurisdictions, with around 90\% coming from US, EP, WO, JP, or CN. In total, we utilize approximately 31.7M citations formed from around 8.7M applications and 14.2M cited documents. 

Each citation is represented by a pair of graphs, where the citing graph is typically the first claim graph, and the cited graph is the description graph. 
At training time, we use the following data augmentations and regularization techniques:
\begin{enumerate}
    \item \emph{Trivial citations:} We create artificial citations from the first claim graph and the description graph of the same patent. We observe that having such citations stabilizes the training.
    \item \emph{Graph type augmentation:} For each sample, with a predefined probability (0.4 in our experiments), we randomly replace either the citing graph or the cited graph (but not both) with the all claims graph. In addition, a replacement of the cited graph is forced for the easiest training samples according to one of our previous models.
    \item \emph{Node dropout:} We randomly drop some nodes from the graphs completely 
    with probabilities that depend on the size of the graph.
    \item \emph{Embedding dropout:} Token embeddings retrieved from the embedding layer (Section \ref{sec:embedding-layer}) are subject to regular dropout.
\end{enumerate}

Non-intersection between training, validation, and test sets is achieved on the document level, and no citations cross the different sets. When excluding a document from a set, we exclude all the documents of the same patent family (i.e., other applications corresponding to the same invention).



\subsection{Training Procedure} \label{sec:training-procedure}

We train our model using the PyTorch \cite{pytorch} and DGL \cite{dgl} libraries. We employ triplet loss~\cite{schroff2015facenet} as our loss function, assigning different margins to different citation categories. We use the AdamW~\cite{loshchilov2017decoupled} optimizer, reducing the learning rate by two on plateaus. Negatives are obtained using online hard negative mining~\cite{shrivastava2016training} over the current batch. Batch creation accounts for graph sizes, making batch size dynamic, with 2100--2260 anchors and 900--960 positives on average. To make batches harder, we group samples into batches using International Patent Classification (IPC)~\cite{WIPO_IPC} classes.

The training has two stages: A \textit{base stage} with an output dimension of 2048, and a \textit{dimensionality reduction stage} reducing the output dimension to 150 by fine-tuning the base stage and adding the MoE layer described in Section \ref{sec:dimensionality-reduction}. While the base stage achieves higher recall, the dimensionality reduction stage strikes a better balance in terms of performance and vector storage cost.

The model is evaluated thrice per epoch.
The training of each stage stops when the top-3 X citation recall~(Recall@3, see Section~\ref{sec:evaluation}) does not improve for three subsequent evaluation runs. When trained on eight L4 GPUs, the training process for both stages combined required approximately 185k updates (12 epochs) and took about 4.6 days.

\begin{table*}[t]
    \centering
    \caption{Performance on retrieval of novelty-destroying citations. Higher is better for both recall and nDCG.}
    \label{tab:results}
    \begin{tabular}{lrrrrr}
        \toprule
        Approach & Seq. len. & Output dim. & Model size & Recall@3 & nDCG@150 \\
        \midrule\multicolumn{6}{c}{\emph{Dense retrieval models trained to embed patent documents}}\\
        \midrule
        Our, base stage & N/A & 2048 & 156M & \textbf{0.4046} & \textbf{0.5564}\\
        Our, dimensionality reduction stage & N/A & 150 & 161M & 0.3861 & 0.5372\\
        Tree-LSTM, base stage & N/A & 600 & 20M & 0.3151 & 0.4685\\
        PaECTER (\texttt{mpi-inno-comp/paecter}) & 4$\times$512 & 1024 & 345M & 0.2798 & 0.4341\\
        \midrule\multicolumn{6}{c}{\emph{Dense retrieval models trained to embed text in general}}\\
        \midrule
        Stella (\texttt{NovaSearch/stella\_en\_400M\_v5}) & 3$\times$2048 & 1024 & 435M & 0.2734 & 0.4134\\
        KaLM (\texttt{HIT-TMG/KaLM-embedding-multilingual-mini-v1}) & 1$\times$4096 & 896 & 494M & 0.2211 & 0.3527\\
        GTE-ModernBert (\texttt{Alibaba-NLP/gte-modernbert-base}) & 1$\times$4096 & 768 & 149M & 0.2003 & 0.3231\\
        \midrule\multicolumn{6}{c}{\emph{Sparse retrieval models}}\\
        \midrule
        Okapi BM25 ($k_1 = 2.7, b = 1.15$) & N/A & N/A & N/A & 0.1866 & 0.2874\\
        \bottomrule
    \end{tabular}
\label{tab:comparisons}
\end{table*}

\section{Results}

To evaluate the performance of our model in patent retrieval, we measure its effectiveness in retrieving novelty-destroying (X) citations of patent applications, as described in Section \ref{sec:evaluation}. The query is the first independent claim of the application, while we use the full text (all claims and description) for the search candidate documents. For our Graph Transformer model, we use the graphs corresponding to the query and search candidate documents as input, and the other approaches use the original texts. 
Any non-English documents are machine translated into English before being processed.

We compare our model with four text embedding models (Section~\ref{sec:comparison-transformers}), our previous Tree-LSTM-based approach (Section~\ref{sec:comparison-tree-lstm}) and Okapi BM25 (Section~\ref{sec:comparison-bm25}).

\subsection{Evaluation Procedure}
\label{sec:evaluation}

To perform the evaluation, we use a test set containing about 161,000 search candidate documents and around 96,000 queries that cite one of the candidate documents as an X citation. We populate the search space by embedding each search candidate into a single vector, i.e., we do document retrieval, not passage retrieval. For each query we then perform a nearest neighbor search among all the search candidate documents and measure how often, on average, the document cited as X appears in the top three results (Recall@3). As an auxiliary metric, we use nDCG~\cite{jarvelin2002cumulated}, which scores relevant document hits based on their positions. To compute nDCG, we use the top 150 search results (nDCG@150). 

\subsection{Comparison with Text Embedding Models}
\label{sec:comparison-transformers}

Based on MTEB leaderboard~\cite{mteb_leaderboard} scores, we selected three publicly available models to evaluate: Stella~\cite{zhang2024jasper}, KaLM~\cite{hu2025kalm} and GTE-ModernBert~\cite{zhang2024mgte}. We selected these instead of larger models since they are of similar size as our model. We also compare our approach with PaECTER~\cite{ghosh2024paecter} as it is specifically trained using patent data. 

We tuned the sequence length of these models to achieve the best Recall@3 on the validation set and then evaluated the models on the test set. To preserve more text, the two models that performed best on our evaluation sets (PaECTER and Stella) were applied on several chunks of the input text and averaged the embeddings. 

PaECTER was trained on patent titles concatenated with abstracts as both queries and search candidates. However, we found that using the text of first claims as queries does not affect model performance, while using the full text (truncated to fit the 512 token window) for candidates even improves it. Thus, the only difference in inputs to PaECTER compared to other text embedding models is prepending input texts with patent titles. 

Table~\ref{tab:comparisons} shows the chosen parameters of all models (resultant sequence lengths and the maximum number of input chunks, which is indicated by a digit before the ``$\times$'' symbol), model sizes and the obtained metric values.
The results show that our Graph Transformer-based approach outperforms the other models on novelty-destroying patent document retrieval. 
The recall of PaECTER is about 31\% lower than our base stage model, while PaECTER performs about as well as Stella and better than the other text embedding models.

\subsection{Comparison with Tree-LSTM}
\label{sec:comparison-tree-lstm}

We also compare our approach with our previous approach based on Tree-LSTM~\cite{bjorkqvist2023building}, which was re-trained using the same training data used to train the Graph Transformer.
Table~\ref{tab:comparisons} shows that its recall is 22\% worse than our Graph Transformer base stage model, while outperforming PaECTER and all other evaluated text embedding models.

\subsection{Comparison with BM25}
\label{sec:comparison-bm25}

BM25~\cite{robertson1995okapi} is a popular traditional word counting-based information retrieval approach that does not use machine learning. We tuned the hyperparameters of the Okapi BM25 version of this approach ($k_1$ and $b$) on 1,000 citations from our validation set (but preserving all available search candidates) and then computed its metrics on the test set. As seen in Table~\ref{tab:comparisons}, the recall of BM25 is less than half of that of our base stage model and also trails all text embedding models we evaluated.

\subsection{Limitations of Comparisons}
\label{sec:comparison-threats}

Our comparison in Section~\ref{sec:comparison-transformers} is limited since, apart from PaECTER, the other models were not trained specifically for patent retrieval. On the other hand, we found that using a large batch size is crucial for high recall when using online hard negative mining. If we would fine-tune text embedding models we couldn't use nearly as large batches as with our approach. With a sequence length of 512 and the GPU setup used for training our model, we could fit around 45-65 positives in a batch (depending on the model), which is more than 13 times less than we achieve with our Graph Transformer approach. This would also make the training process significantly slower and could impact the results due to the truncated texts.

We also note that this evaluation focuses on first-stage retrieval efficiency and effectiveness; comparison with computationally intensive re-ranking models like cross-encoders was considered beyond the scope of this specific study but represents a potential area for future investigation.


\section{Conclusion}

In this paper, we presented an approach for patent searching that utilizes a Graph Transformer model as a dense retriever. We compared the performance of our approach to existing text embedding models and showed that our approach achieves a significantly higher recall on novelty-destroying citation retrieval. Additionally, we demonstrated that our approach is more computationally efficient than existing text-based Transformer models.

\bibliography{sample-base}

\end{document}